\documentclass[a4paper,fleqn,usenatbib]{mnras}

\usepackage{newtxtext,newtxmath}
\usepackage[T1]{fontenc}
\usepackage{ae,aecompl}

\usepackage{graphicx}	% Including figure files
\usepackage{amsmath}	% Advanced maths commands
\usepackage{amssymb}	% Extra maths symbols

\usepackage{ulem}
\usepackage{xcolor}
\usepackage{soul}
\providecolor{added}{rgb}{1,0,0}
\providecolor{deleted}{rgb}{0,0,1}

\title[Methanol in the disc of TW~Hya]{On the methanol emission detection in the TW Hya disc: the role of grain surface chemistry and non-LTE excitation}

\author[S. Yu. Parfenov et al.]{
S. Yu. Parfenov,$^{1}$\thanks{E-mail: sergey.parfenov@urfu.ru}
D. A. Semenov,$^{2}$
Th. Henning,$^{2}$
A. S. Shapovalova,$^{1}$
A. M. Sobolev,$^{1}$
\newauthor R. Teague$^{2}$
\\
$^{1}$Ural Federal University, 51 Lenin Str., Ekaterinburg 620000, Russia\\
$^{2}$Max Planck Institute for Astronomy, K{\"{o}}nigstuhl 17, D-69117 Heidelberg, Germany
}

\date{Accepted XXX. Received YYY; in original form ZZZ}

\pubyear{2017}

\begin{document}
\label{firstpage}
\pagerange{\pageref{firstpage}--\pageref{lastpage}}
\maketitle

\begin{abstract}
The recent detection of gas-phase methanol (CH$_3$OH) lines in the disc of TW~Hya by Walsh et al. provided the first observational constraints on the complex O-bearing organic content in protoplanetary discs. The emission has a ring-like morphology, with a peak at $\sim 30-50$~au and an inferred column density of $\sim 3$--$6\times10^{12}$~cm$^{-2}$. A low CH$_3$OH fractional abundance of $\sim 0.3$--$4\times 10^{-11}$ (with respect to H$_2$) is derived, depending on the assumed vertical location of the CH$_3$OH molecular layer. In this study, we use a thermo-chemical model of the TW Hya disc, coupled with the \textsc{alchemic} gas-grain chemical model, assuming laboratory-motivated, fast diffusivities of the surface molecules to interpret the CH$_3$OH detection. Based on this disc model, we performed radiative transfer calculations with the \textsc{lime} code and simulations of the observations with the \textsc{casa} simulator. We found that our model allows to reproduce the observations well. The CH$_3$OH emission in our model appears as a ring with radius of $\sim60$~au. Synthetic and observed line flux densities are equal within the rms noise level of observations. The synthetic CH$_3$OH spectra calculated assuming local thermodynamic equilibrium (LTE) can differ by up to a factor of 3.5 from the non-LTE spectra. For the strongest lines, the differences between LTE and non-LTE flux densities are very small and practically negligible. Variations in the diffusivity of the surface molecules can lead to variations of the CH$_3$OH abundance and, therefore, line flux densities by an order of magnitude.
\end{abstract}


%%%%%%%%%%%%%%%%%%%%%%%%%%%%%%%%%%%%%%%%%%%%%%%%%%

%%%%%%%%%%%%%%%%% BODY OF PAPER %%%%%%%%%%%%%%%%%%

\section{Introduction}

The architecture of planetary systems and the composition of extrasolar planets appear to be tightly linked with the structure and chemical evolution of their birth sites, protoplanetary discs \citep{Mordasini2012a,Mordasini2012b,Winn2015}.
The disc chemical composition may additionally shape the properties of exoplanetary atmospheres \citep[e.g.][]{Bergin_ea15,Cridland2016,Mordasini_ea2016}. Furthermore, some of complex organic (pre-biotic) molecules in the primordial planetary atmospheres may be formed during the natal disc evolution \citep[see e.g.][]{Ciesla2012,Caselli2012,Henning2013}. The results from the recent {\sc Rosetta} mission to the comet 67P/Churyumov-Gerasimenko indicate that highly complex organic molecules, including methanol, were present at the verge of planet formation in the solar nebula  and could have been later delivered on early Earth during the Late Heavy Bombardement phase \citep{Biver2015,organics_Rosetta_2016}.

So far, however, there were only a few organic species with O- or N-atoms have been detected in discs, namely, H$_2$CO \citep{Aikawa2003}, HCN, HNC, HC$_3$N \citep{Chapillon2012}, and CH$_3$CN \citep{Oberg2015}. Despite intense searches, methanol emission was not detected in discs until recently \citep[][hereafter W16]{Walsh2016}. CH$_3$OH is mainly synthesized via surface chemistry associated with CO and plays an important role in the formation of more complex organic species \citep[see e.g.][]{Herbst2009}. The study of \citet{Drozdovskaya2014} showed that the methanol ice in the outer disc regions may be inherited from the parental clouds in the prestellar stage. These clouds have detectable amounts of the methanol both in gas phase \citet{Kalenskii1994} and on the grain surface \citet{Gurtler2002}. The capacity of the methanol reservoir within the grain mantles can be demonstrated by the fact that release of the methanol from the grain mantles can increase the gas-phase abundance of this molecule, $X_{\rm{M}}$ (wrt H$_2$), by many orders of magnitude \citet{Sutton2004}. The methanol gas-phase abundance predicted by different chemical disc models varies by orders of magnitude \citep[see e.g.][]{Walsh2014}, with the main source of uncertainty in $X_{\rm{M}}$ being due to the poorly known chemical reaction rates \citep[see e.g.][]{Vasyunin2008} and the binding and desorption energies of ices. 
Furthermore, the CH$_3$OH lines in discs can be subject to non-local thermodynamic equilibrium (non-LTE) excitation and thus a full non-LTE line radiative transfer (LRT) modelling is necessary \citep{Parfenov2016}.

The detection of CH$_3$OH emission in the disc of TW~Hya by W16 is a great achievement in the characterization of the chemical complexity in protoplanetary discs. W16 estimated $X_{\rm{M}}\sim0.3$--$4\times10^{-11}$ utilising the disc physical model of \citet{Kama2016} and a parametrized CH$_3$OH spatial distribution. This estimate is lower by two orders of magnitude than those predicted by contemporary chemical models of T~Tauri discs. Such an inconsistency can be related to both the incompleteness of chemical models and the differing disc structure of TW Hya to the discs structures typically adopted for disc chemical modelling \citep[by e.g.][]{Willacy2007,SW2011,Walsh2014,Furuya2014}.

In this paper, we continue our study of complex organic chemistry and line excitation in discs. We interpret the CH$_3$OH emission detection using a physical model of the TW~Hya disc coupled with a complex chemical model (Section~\ref{sec:model}). The CH$_3$OH distribution computed with this disc model is used for non-LTE and LTE LRT calculations (Section~\ref{sec:lrt}). Based on the results of LRT calculations we simulate observations of the TW~Hya disc with the Atacama Large Millimeter/submillimeter Array (ALMA) and compare these simulated observations with W16 results (Section~\ref{sec:obs}). A discussion and conclusions follow.

\section{TW Hya disc model}
\label{sec:model}

For the physical structure of the TW~Hya disc, we adopted the thermo-chemical model from \citet{Gorti2011}. This model is based on the iterative fitting of multiple gas emission spectra and dust emission data ranging from the sub-millimetre to the optical wavelengths. The dust model is taken from \citet{Calvet2002} and has a total dust mass of $2.4\times 10^{-4}\,\rm{M}_{\rm{\sun}}$, following a grain size distribution $n(a) \propto a^{-3.5}$ with a maximum grain size $a_{\rm max}=1$~mm. The assumed polycyclic aromatic hydrocarbons abundance per H atom is $\sim 10^{-9}$. A constant dust-to-gas mass ratio of 100 is assumed. An accretion rate of $10^{-9}\,\rm{M}_{\rm{\sun}}$~yr$^{-1}$ is assumed.
The disc model spans $3.9$--$200$~au radially and has separately computed gas and dust temperatures. To compute the gas thermal balance, a concise chemistry model with $\sim 85$ species and $\sim 600$ reactions is utilized.

The adopted parameters of the central star and the incident ultraviolet (UV) and X-ray radiation fields represent those observed in TW~Hya. The star has a mass of $0.7\,\rm{M}_{\rm{\sun}}$, a radius of $1\,\rm{R}_{\rm{\sun}}$, and an effective temperature of 4200~K. A detailed far-UV spectrum with a total far-UV luminosity of $3\times10^{31}$~erg\,s$^{-1}$ was used. The adopted X-ray spectrum, covering $0.1$--$10$~keV, has a total X-ray luminosity of $1.6\times10^{30}$~erg\,s$^{-1}$.

The chemical structure of the disc was computed with the gas-grain code \textsc{alchemic} described in detail by \citet{Semenov2010,SW2011} and used in our previous methanol study, \citet{Parfenov2016}. A new lower ultraviolet-photodesorption yield of $10^{-5}$ is adopted for all species, based on the recent measurements of \citet{Cruz_Diaz2016} and \citet{Bertin2016}. While the inferred age of the TW~Hya system is about 7~Myr \citep{Ducourant2014}, we modelled the disc chemical evolution over only 1~Myr. This was dictated by the slow performance of our gas-grain code, limited by the very slow pace of surface processes on $7\mu$m-sized grains. We note that CH$_3$OH abundances are in a quasi-steady state at 0.2--1~Myr in the disc regions with high CH$_3$OH abundances ($X_{\rm{M}}\ga 10^{-12}$).

\begin{figure*}
	\includegraphics[scale=0.44]{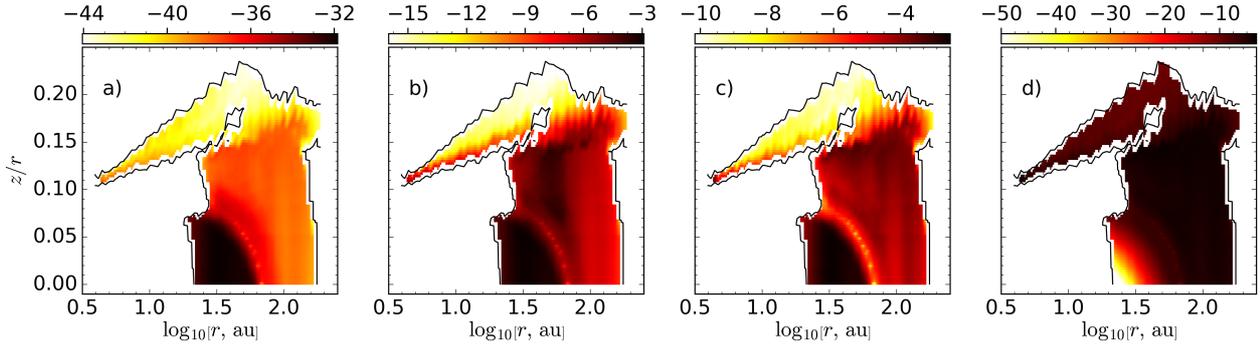}
    \caption{The log of methanol desorption rates in cm$^{-3}$~s$^{-1}$ estimated for the `$E_{\rm{diff}}/E_{\rm{b}}=0.4$' disc model. The rates are shown only in the disc regions with $X_{\rm{M}}>10^{-15}$. (a) CRP desorption rate. (b) Reactive desorption rate. (c) Photodesorption rate. (d) Photodesorption rate estimated without taking into account the secondary UV photons produced by CRP excitation of H$_2$. The black contours denote the gas-phase CH$_3$OH abundance of $10^{-15}$.}
    \label{fig:rates}
\end{figure*}

In our model, CH$_3$OH is formed on dust surfaces via the hydrogenation of CO, in which hydrogen tunnelling through reaction barriers is taken into account, however tunnelling through potential wells of surface sites is neglected. Additionally, CH$_3$OH can be formed on dust surfaces via radical-radical recombination between OH and CH$_3$, which in turn are formed by surface recombination reactions and UV photoprocessing of more complex `parent' molecules. CH$_3$OH can also be formed inefficiently in the gas phase via dissociative recombination of CH$_3$OH$_2^+$ with $e^-$, which, in our network, has a 5 per cent probability to produce CH$_3$OH and H. The methanol in our model is released into the gas-phase mainly through chemical desorption and photodesorption of the CH$_3$OH ice by UV radiation (see Fig.~\ref{fig:rates}). Both the interstellar UV and the secondary UV radiation produced by the cosmic ray particle (CRP) excitation of H$_2$ are important. The peak temperature of the grain particle due to the impulsive heating from CRPs is estimated taking into account an initial grain temperature with equation (6) from \citet{Semenov2004} and is relatively low comparing with methanol desorption energy (5530~K). Thus, CRP induced desorption through the grain heating is ineffective in our model. Chemical desorption, with an efficiency of 1 per cent, occurs for all surface recombination reactions, including CH$_3$OH formation.

The values of $X_{\rm{M}}$ used in \citet{Parfenov2016} were computed for a solar nebula model from \citet{SW2011} and are lower by about two orders of magnitude compared to the chemical model of similar complexity from \citet{Walsh2014}. As noted in \citet{Parfenov2016}, this discrepancy can be due to the relatively low grain-surface diffusion rates assumed by \citet{SW2011}, with the ratio of diffusion to binding energy $E_{\rm{diff}}/E_{\rm{b}}$ of 0.77 for surface species. In this study, however, we consider surface reactions with faster diffusion rates with $E_{\rm{diff}}/E_{\rm{b}}=0.4$, close to the $E_{\rm{diff}}/E_{\rm{b}}=0.3$ used by \citet{Walsh2014}. A value of $E_{\rm{diff}}/E_{\rm{b}}=0.4$ is more consistent with the recent laboratory studies of \citet{Cuppen2017}. In order to see how the disc chemical composition can be affected by the uncertainties of $E_{\rm{diff}}/E_{\rm{b}}$ we also performed calculations with $E_{\rm{diff}}/E_{\rm{b}}=0.77$. Our models with $E_{\rm{diff}}/E_{\rm{b}}=0.77$ and $0.4$ will be denoted as `$E_{\rm{diff}}/E_{\rm{b}}=0.77$' and `$E_{\rm{diff}}/E_{\rm{b}}=0.4$' models, respectively.

\begin{figure*}
	\includegraphics[scale=0.44]{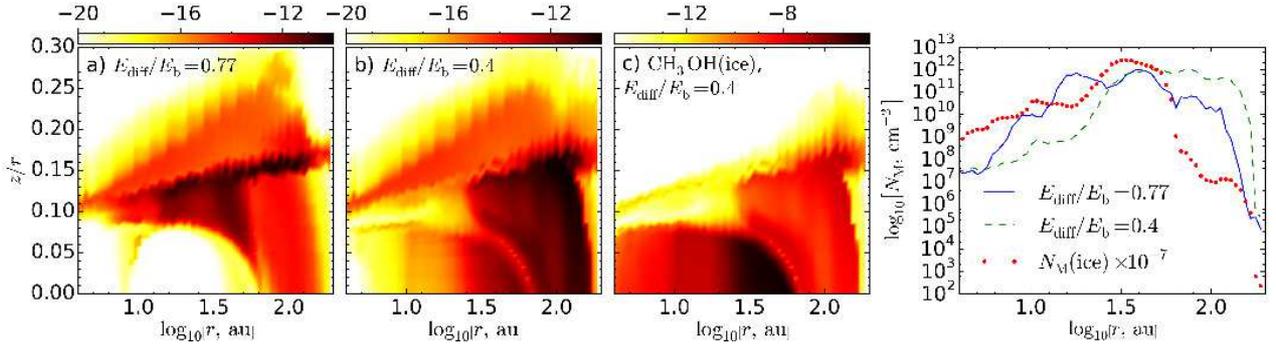}
    \caption{(a) and (b) The log of relative gas-phase CH$_3$OH abundance with respect to the H$_2$ density in the `$E_{\rm{diff}}/E_{\rm{b}}=0.77$' and `$E_{\rm{diff}}/E_{\rm{b}}=0.4$' disc models. (c) Same as (b) but for grain surface (ice) CH$_3$OH. The most right panel shows gas-phase CH$_3$OH vertical column density in the `$E_{\rm{diff}}/E_{\rm{b}}=0.77$' (solid line) and `$E_{\rm{diff}}/E_{\rm{b}}=0.4$' (dashed line) chemical models and grain-surface (ice) CH$_3$OH vertical column density in the `$E_{\rm{diff}}/E_{\rm{b}}=0.4$' model divided by $10^7$ (dotted line).}
    \label{fig:Xm}
\end{figure*}

In both `$E_{\rm{diff}}/E_{\rm{b}}=0.77$' and `$E_{\rm{diff}}/E_{\rm{b}}=0.4$' chemical models, gas-phase CH$_3$OH resides mainly in the disc regions with $z/r<0.25$ (see Fig.~\ref{fig:Xm}), where $z$ denotes the height above the disc mid-plane and $r$ is the distance from the disc centre. This is in good agreement with the spatial distribution of CH$_3$OH considered by W16. The maximum CH$_3$OH abundance, $X_{\rm{0M}}$, in `$E_{\rm{diff}}/E_{\rm{b}}=0.77$' and `$E_{\rm{diff}}/E_{\rm{b}}=0.4$' models is comparable: $1.87\times10^{-10}$ and $1.55\times10^{-10}$, respectively. However, the gas-phase CH$_3$OH in the `$E_{\rm{diff}}/E_{\rm{b}}=0.4$' model occupies a much larger region of the disc, including the disc mid-plane, in contrast to the `$E_{\rm{diff}}/E_{\rm{b}}=0.77$' model. This leads to a difference of several orders of magnitude in CH$_3$OH vertical column densities, $N_{\rm{M}}$, between `$E_{\rm{diff}}/E_{\rm{b}}=0.77$' and `$E_{\rm{diff}}/E_{\rm{b}}=0.4$' models for the disc regions with $r<20$ and $r>50$~au (see Fig.~\ref{fig:Xm}). For example, $N_{\rm{M}}=9\times10^{10}$ and $7.1\times10^{11}$~cm$^{-2}$ at $r=60$~au for `$E_{\rm{diff}}/E_{\rm{b}}=0.77$' and `$E_{\rm{diff}}/E_{\rm{b}}=0.4$' models, respectively.

\begin{figure*}
	\includegraphics[scale=0.3,,trim={0 0.9cm 0 0.5cm}]{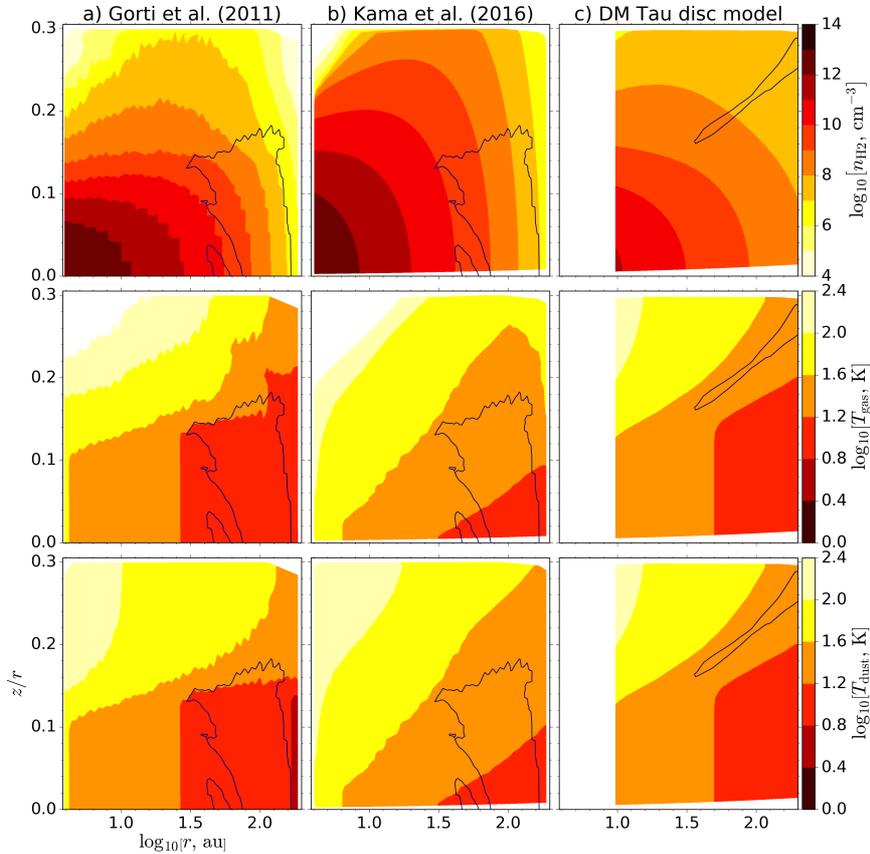}
    \caption{TW Hya and DM Tau disc models. Left and middle columns show relative methanol abundances (wrt H$_{2}$) computed with the physical models of \citet{Gorti2011} and \citet{Kama2016}, respectively. Right column shows the methanol abundances computed with the DM~Tau disc model taken from \citet{SW2011}. The black contours denote the methanol abundance of $10^{-12}$.}
    \label{fig:phys_models}
\end{figure*}

The disc physical structure used in this study is different from the one of \citet{Kama2016} adopted by W16. As can be clearly seen in Figure~\ref{fig:phys_models}, the Kama et al. disc is denser and colder compared to the disc model of \citet{Gorti2011}. To assess how the differences between adopted disc physical models affect the predicted CH$_3$OH line flux densities, we performed LRT calculations with our disc physical model but using the parametrized CH$_3$OH distribution considered by W16. It is worth noting that the \citet{Gorti2011} disc model does not have separated mm- and $\mu$m-sized grain populations \citep[see e.g.][]{Menu2014}. Both the \citet{Gorti2011} and \citet{Kama2016} disc models do not take into account the gaps seen in scattered light and in the ALMA (sub-)mm images \citep[see e.g.][]{Debes2013,vanBoekel2016,Andrews2016}. These gaps can be associated with the deficit of large ($\sim10$~mm) dust grains \citep{Tsukagoshi2016} or a local gas density depletion \citep{Teague2017}.

\section{Radiative transfer calculations}
\label{sec:lrt}

For the LRT calculations, we have used the LIne Modelling Engine (\textsc{lime}) \citep{Brinch2010} and the `$v_{\rm{t}}=0$' CH$_3$OH level scheme of \citet{Parfenov2016} that is very similar to the one given in the Leiden Atomic and Molecular Database
\citep[LAMDA,][]{Schoier2005}. We assumed that $A$ and $E$ species of CH$_3$OH are equally abundant.

We used the dust opacities from \citet{Menu2014} for our LRT calculations. As W16 do not state explicitly which dust opacities they used, it makes a comparison of the results between W16 LRT calculations and our test calculations with the W16 CH$_3$OH parametrized distribution more difficult. To check how the variations in dust opacities can affect the results of our test LRT calculations we performed these calculations using dust opacities from both \citet{Menu2014} and \citet{Andrews2011}. The latter dust opacities are consistent with those utilized in \textsc{dali} code \citep{Bruderer2013}, which in turn have been used by \citet{Kama2016} to construct the model of TW Hya disc. At frequencies of $\sim305$ GHz, \citet{Menu2014} dust opacity is higher by about a factor of $\sim1.8$ than the one of \citet{Andrews2011}.

The spatial grid of $1\,000\,000$ points used for LRT calculation consisted of 5 subsets of points, each including $200\,000$ points. The positions of points in three of these subsets were selected from weighted random distributions with weighting functions $\left( n_{\rm{H2}}/n_{\rm{0H2}} \right)^{0.3}$, $\left( X_{\rm{M}}/X_{\rm{0M}} \right)^{0.3}$ and $\left[ (X_{\rm{M}}n_{\rm{H2}})/(X_{\rm{0M}}n_{\rm{0H2}}) \right]^{0.3}$, where $n_{\rm{0H2}}=4.2\times10^{12}$~cm$^{-3}$ is the maximum value of H$_2$ density, $n_{\rm{H2}}$. These 3 subsets trace the disc regions with high gas and CH$_3$OH densities. The remaining 2 subsets were constructed to trace the disc regions with relatively low gas density and $X_{\rm{M}}$. The positions of points in these 2 subsets were sampled from random distribution that are uniform in Cartesian coordinates ($x$,$y$,$z$) and in $\log_{10}\left( \sqrt{x^2+y^2+z^2} \right)$, respectively.

The synthetic images in CH$_3$OH lines were obtained with a pixel size of $0.01$~arcsec, which allows us to well resolve the $\sim1\arcsec\times1\arcsec$ beam, comparable to the one in W16. As in W16, the synthetic images were obtained with a spectral resolution of $0.15$~km~s$^{-1}$, assuming a disc inclination and position angle of $7$ and $155\degr$ \citep{Qi2004}, respectively. We adopted a distance to the TW Hya disc of 59.5 pc from the first \textit{Gaia} data release \citep{Gaia}. For consistency with W16, the test calculations with the \citet{Gorti2011} disc model and parametrized CH$_3$OH distribution of W16 were performed assuming a distance of 54 pc. By adding line specific intensities in the all pixels of the synthetic images, we obtained the synthetic integrated CH$_3$OH spectra.

\textsc{lime} produces images at a single spectral frequency without taking into account the averaging within the spectral channels as in real observations. We have checked how this averaging affects the synthetic spectra following the procedure given by \citet{Cleeves2016}. We performed calculations of the synthetic spectra with a spectral resolution of $0.015$~km~s$^{-1}$, ten times higher than our initial resolution and the resolution of the W16 data, before averaging the results of these computations back to the resolution of $0.15$~km~s$^{-1}$. The resulting synthetic peak line flux densities do not vary by more than 3 per cent due to the averaging, a variation which is significantly lower than the ratio of the rms noise (6.5 mJy) to the peak flux density (33 mJy) in the spectrum of W16. Thus we assume the averaging can safely be neglected.

\begin{figure}
	\includegraphics[width=\columnwidth{},trim={0 0.5cm 0 6cm},clip]{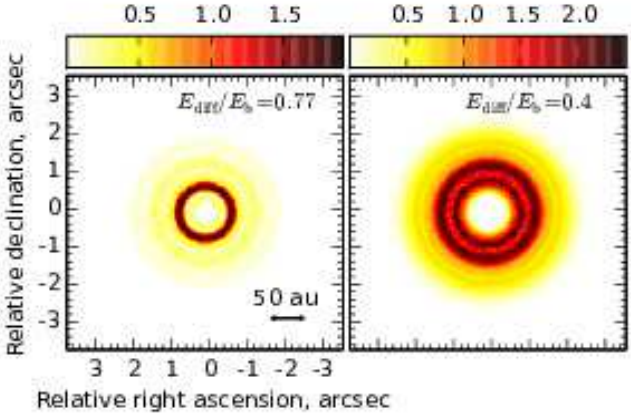}
    \caption{The zeroth moment maps obtained for the $3_1-3_0A^{-+}$ (305.473~GHz) transition with the `$E_{\rm{diff}}/E_{\rm{b}}=0.77$' (left panel) and `$E_{\rm{diff}}/E_{\rm{b}}=0.4$' (right panel) disc chemical models. The colour bars denote the intensity in $10^{-6}$ Jy~pixel$^{-1}$~km~s$^{-1}$.}
    \label{fig:zero_mom}
\end{figure}

W16 considered the Band 7 $2_1-2_0A^{-+}$ (304.208~GHz), $3_1-3_0A^{-+}$ (305.473~GHz), $4_1-4_0A^{-+}$ (307.166~GHz) transitions (hereafter, B7 lines) and Band 6 $5_0-4_0E$ (241.700~GHz), $5_{-1}-4_{-1}E$ (241.767~GHz), $5_0-4_0$ (241.791~GHz) transitions (hereafter, B6 lines). The difference in the maximum specific intensity of the synthetic images for these lines between the `$E_{\rm{diff}}/E_{\rm{b}}=0.77$' and `$E_{\rm{diff}}/E_{\rm{b}}=0.4$' models does not exceed 45 per cent. The spatial distribution of the CH$_3$OH emission, however, is significantly different, primarily caused by the difference in the methanol column densities $N_{\rm{M}}$ between the models (see Sect.~\ref{sec:model}). In the case of the `$E_{\rm{diff}}/E_{\rm{b}}=0.77$' model, the CH$_3$OH emission is mainly concentrated within a thin ring, while in the case of the `$E_{\rm{diff}}/E_{\rm{b}}=0.4$' model it is distributed over a larger image area (see Fig.~\ref{fig:zero_mom}).

\begin{figure}
	\includegraphics[width=\columnwidth{},trim={0 0.5cm 0 1cm}]{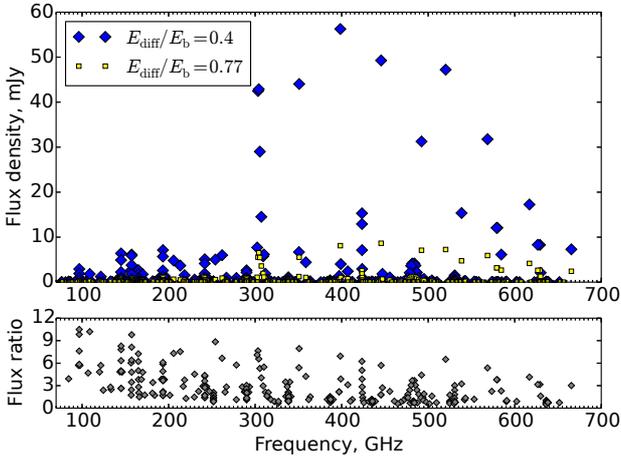}
    \caption{The synthetic spectra computed with the \textsc{lime} code. Upper panel shows the line peak flux densities of CH$_3$OH calculated for the `$E_{\rm{diff}}/E_{\rm{b}}=0.77$' and `$E_{\rm{diff}}/E_{\rm{b}}=0.4$' TW~Hya disc chemical models. Lower panel shows the ratio of line peak flux densities in the `$E_{\rm{diff}}/E_{\rm{b}}=0.77$' and `$E_{\rm{diff}}/E_{\rm{b}}=0.4$' models.}
    \label{fig:sw11w14}
\end{figure}

The line flux densities, $F_{\nu}$, in the spectra obtained with the `$E_{\rm{diff}}/E_{\rm{b}}=0.77$' model
are lower by a factor of $<9$ than those obtained with the `$E_{\rm{diff}}/E_{\rm{b}}=0.4$' model (see Fig.~\ref{fig:sw11w14}). The difference in $F_{\nu}$ between the two chemical models increases with increasing $F_{\nu}$ and with decreasing transition frequency.

\begin{figure}
	\includegraphics[width=\columnwidth{},trim={0 0.5cm 0 1cm}]{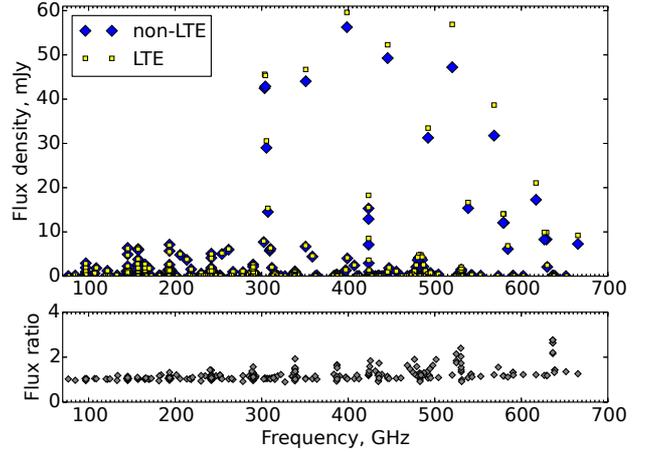}
    \caption{The synthetic spectra calculated with \textsc{lime} code for the `$E_{\rm{diff}}/E_{\rm{b}}=0.4$' TW~Hya disc chemical model. Upper panel shows the line peak flux densities of CH$_3$OH in LTE and non-LTE synthetic spectra. Lower panel shows the ratio of LTE and non-LTE line peak flux densities. Only lines with peak flux densities $>0.01$~mJy are shown.}
    \label{fig:specs}
\end{figure}

The synthetic LTE CH$_3$OH spectra do not differ by more than a factor of $3.5$ from the non-LTE spectra for both the `$E_{\rm{diff}}/E_{\rm{b}}=0.77$' and `$E_{\rm{diff}}/E_{\rm{b}}=0.4$' models (see Fig.~\ref{fig:specs}). For the strongest lines with $F_{\nu}>10$~mJy, the LTE fluxes do not exceed the non-LTE fluxes by more than 30 per cent, while the largest deviations from LTE occur for the $J_{2}\rightarrow{}J_{1}\,A$ and $J_{-2}\rightarrow{}J_{-1}\,E$ line series. The emission distribution in the synthetic images is similar in both LTE and non-LTE cases. This is in contrast to our previous study \citep{Parfenov2016}, where we used the DM ~Tau disc model, and where the majority of methanol lines were found to be sub-thermally excited and had intensities well below their LTE intensities. This discrepancy is due to the adopted disc density structures, where the current model of the TW Hya disc is denser than the previous DM Tau model (see Figure~\ref{fig:phys_models}). Moreover, in the case of the DM Tau disc, gas-phase methanol resides mainly in the upper disc layers with lower densities compared to the TW Hya disc. Higher gas densities in the TW~Hya model lead to the thermalization of most of the low-energy methanol transitions.

The peak LTE fluxes in B6 and B7 transitions do not differ from the non-LTE ones by more than $8$ per cent, a factor of a few lower than the ratio of the noise to the signal in the W16 spectrum. Thus, due to relatively small influence of non-LTE effects on the intensity of B6 and B7 lines, the synthetic images used to simulate the W16 observations in Section~\ref{sec:obs} were calculated assuming LTE. To minimize artefacts caused by finite resolution and randomness of the \textsc{lime} spatial grid we averaged images obtained with 10 different grid realizations. These averaged images were used as input models for the simulations of W16 observations in Section~\ref{sec:obs}.

\begin{figure}
	\includegraphics[width=\columnwidth{},trim={0 0.1cm 0 0.5cm},clip]{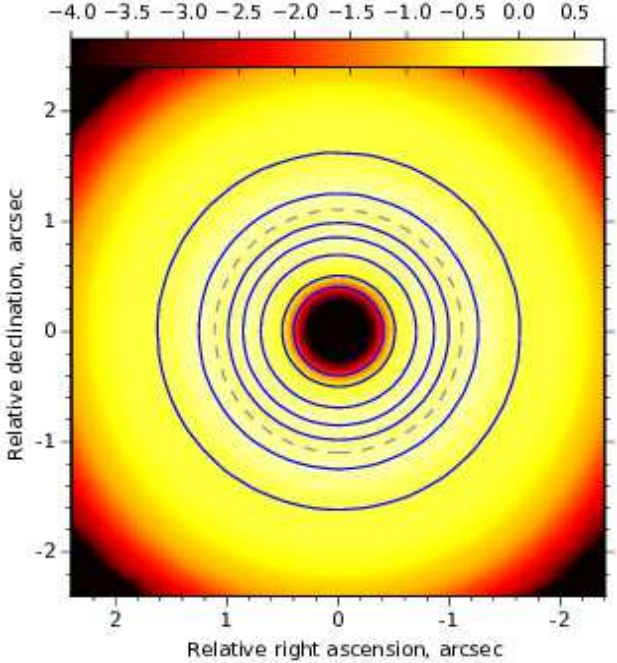}
    \caption{The zeroth moment map obtained for the $3_1-3_0A^{-+}$ (305.473~GHz) transition with the `$E_{\rm{diff}}/E_{\rm{b}}=0.4$' disc chemical model. The colour bar denotes the $\log_{10}$ of intensity in $10^{-6}$ Jy~pixel$^{-1}$~km~s$^{-1}$. Solid contour lines denote the ratio of the maximum CH$_3$OH line specific intensity calculated with zero dust opacity to the line specific intensity calculated with the \citet{Menu2014} dust opacity. The contours from outer to inner are 1.01, 1.1, 1.3, 1.5, 2, 4 and 8. Dashed grey line shows the radius of 66 AU outside which the large grains are depleted according to the (sub-)mm continuum observations of the TW Hya disc by \citep{Andrews2012}.}
    \label{fig:dustOp}
\end{figure}

We have also tested how the optical depth decrease observed in the disc gaps can influence the LRT results. For this purpose, we varied the gas-to-dust ratio keeping the gas density fixed to reproduce the optical depth variations across the gaps estimated by \citet{Tsukagoshi2016}. We have considered the two most prominent gaps at 22 and 37 au from the disc centre and modelled them as Gaussians with a full width at half maximum of 3 au and a depth of 0.25. We did not find any significant influence of the gaps on the modelled disc spectra and images. The spectra obtained with and without gaps do not differ by no more than 8 per cent. The variations of $F_{\nu}$ of B7 and B6 lines due to gaps do not exceed 1 per cent and are thus negligible. It should be noted also that (sub-)mm continuum observations of the TW Hya disc indicate that the spatial distribution of the mm-sized grains in the disc has a sharp outer cutoff \citep{Andrews2012,Hogerheijde2016}. Assuming the distance to TW Hya of 54 pc, \citet{Andrews2012} estimated that this cutoff is located at $r=60$ au which translates to 66 au for the distance of 59.5 pc. The \citet{Gorti2011} and \citet{Kama2016} disc models do not have such a cutoff in their grain populations. Out test LRT calculations with zero dust opacity show that the CH$_3$OH line specific intensities in our model decrease by no more than 25 per cent due to non-zero dust (sub-)mm opacities at $r>66$ au, compared with the case when the dust opacity is absent at $r>66$ au (see Fig.~\ref{fig:dustOp}). The peak flux density increase due to absence of the dust opacity at $r>66$ au does not exceed 5 per cent and thus can be neglected.

\section{Simulation of ALMA observations.}
\label{sec:obs}

The simulations of the TW~Hya disc observations were performed with the \textsc{simobserve} and \textsc{simanalyze} tasks of the \textsc{casa} package version 4.7.0. The parameters of these tasks were chosen to provide the parameters of simulated observations as close as possible to observations of W16. This resulted in synthesized line beams that are $1.\arcsec3\times0.\arcsec8$ ($\rm{PA}=-80\degr$) and $0.\arcsec6\times0.\arcsec5$ ($\rm{PA}=-89\degr$) for the B7 and B6 data, respectively. To perform the uv-stacking of individual transitions within the B7 and B6 datasets we used the same values for the \textsc{simobserve} input frequency of 305.5 and 241.7~GHz, respectively. As in W16, we imaged the line data without CLEANing. The rms noise level in the resulting channel maps for B6 and B7 lines is $\sim4.5$~mJy~beam$^{-1}$. The continuum data at 317~GHz were imaged and CLEANed down to a 0.2~mJy~beam$^{-1}$ threshold using Briggs weighting.

\begin{figure*}
	\includegraphics[scale=1.0,trim={0 1cm 0 0.5cm}]{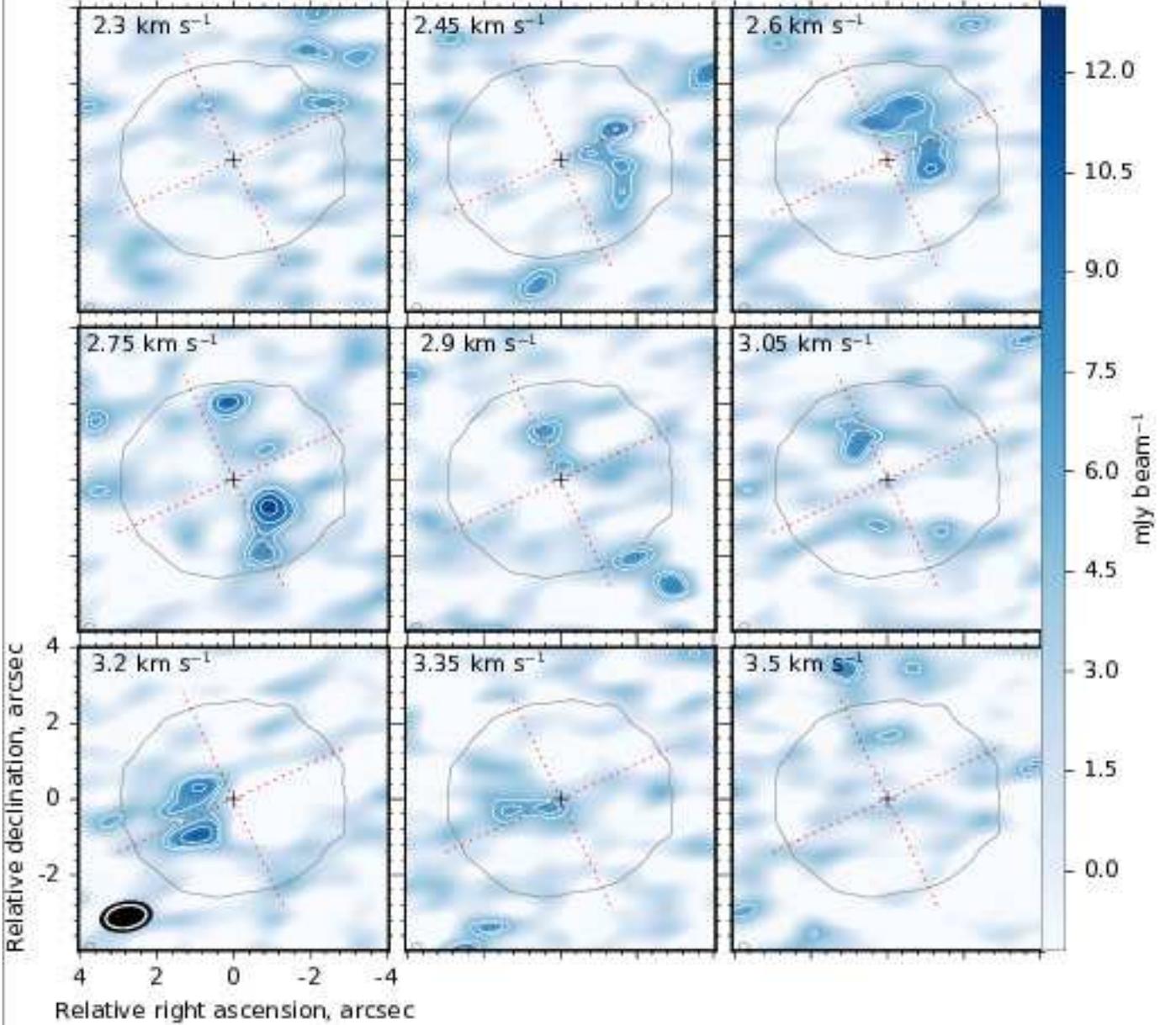}
    \caption{The synthetic channel maps for the stacked B7 data calculated with the `$E_{\rm{diff}}/E_{\rm{b}}=0.4$' TW~Hya disc chemical model. White contours show $2.5\sigma$, $3\sigma$, $4\sigma$ and $5\sigma$ levels for the B7 data. Grey contour shows the $3\sigma$ level of continuum emission at 317~GHz. Dashed lines denote minor and major disc axes and the black cross denotes the stellar position. The synthesized beams for the continuum and line data are shown as open white and filled black ellipses, respectively.}
    \label{fig:chmap}
\end{figure*}

CH$_3$OH emission was not detected in all our channel maps for individual lines except in the channel maps for the $2_1-2_0\,A^{-+}$ transition (304.208~GHz) calculated with the `$E_{\rm{diff}}/E_{\rm{b}}=0.4$' chemical disc model. The emission in this transition was detected at the level of $\gtrsim3\sigma$ in six velocity channels at 2.45, 2.6, 2.75, 3.05, 3.2 and 3.35~km~s$^{-1}$. Following W16, we also performed stacking of visibility data for individual lines within B6 and B7 datasets. We stacked, imaged and CLEANed B6 and B7 data using the \textsc{clean} task with natural weights. An rms noise level in the resulted stacked data for B6 and B7 lines is $\sim2.2$~mJy~beam$^{-1}$. A significant signal is only present in stacked B7 data calculated using the `$E_{\rm{diff}}/E_{\rm{b}}=0.4$' chemical model. The CH$_3$OH emission is detected at the level of $\gtrsim3\sigma$ in seven velocity channels from 2.45 to 3.35~km~s$^{-1}$ (see Fig.~\ref{fig:chmap}).

The methanol emission in our stacked B7 data appears as a ring around the disc centre with a radius of $\sim1.1$~arcsec (60~au). The peak flux in the spectrum obtained by spatial integration of the line emission within $3\sigma$ contour of the 317~GHz continuum is $\sim37$~mJy. Taking into account the rms noise of 6.7 and 6.5 mJy in the spectra of ours and W16 results respectively, the peak integrated flux density of 37~mJy is equal to the peak integrated flux density of $33$ mJy obtained by W16.

The \textsc{lime} images in B7 lines obtained by us using the \citet{Gorti2011} physical model coupled with the best `by-eye' fit chemical model of W16 were stacked by simple averaging. The peak integrated flux density in this stacked data is 28 or 34~mJy depending on whether \citet{Menu2014} or \citet{Andrews2011} dust opacities are utilized, respectively. This is 5~mJy within the synthetic $F_{\nu}=33$ mJy obtained by W16 with the \citet{Kama2016} physical model. Thus the difference in flux densities estimated with the same CH$_3$OH distribution but with the two different yet feasible TW~Hya disc physical models does not exceed the rms noise in both our spectra and the spectra of W16. We conclude that the differences in the disc physical models do not affect significantly the results of our simulations.

\section{Discussion and conclusions}
\label{sec:discus_concl}

The adopted chemical disc model has a complexity comparable to other contemporary disc chemical models. Along with an appropriate TW~Hya disc physical structure this chemical model is able to reproduce the W16 CH$_3$OH detection rather well. Our model predicts a CH$_3$OH abundance that is lower by about two orders of magnitude in comparison with other T~Tauri disc chemical studies with similar models. The reason for such a discrepancy is mainly due to the difference in the adopted disc physical structures rather than the chemical models. Indeed, the TW~Hya disc is more compact and dense compared to the disc models used by e.g. \citet{Willacy2007,SW2011,Walsh2014,Furuya2014}.

The CH$_3$OH abundance on the grain surfaces predicted with our TW~Hya disc model is relatively high and comparable with the estimates of \citet{Walsh2014} (see Fig.~\ref{fig:Xm}). Methanol ice in our TW Hya disc model reaches its highest concentrations at around $30$--$50$ au from the disc centre, where the temperatures in the disc mid-plane are most favourable for CO hydrogenation on grain surfaces. One of the main formation routes of gas-phase methanol in our model is the photodesorption of methanol ice by interstellar UV photons and UV photons produced by CRP excitation of H$_2$. The effectiveness of the UV photodesorption in dense discs is restricted to layers well above the disc mid-plane. In contrast, CRPs and partly attenuated, very hard X-ray photons are able to penetrate larger gas columns and are able to maintain a relatively high gas-phase CH$_3$OH abundance in the molecular layers adjacent to the mid-plane. The maximum of the CH$_3$OH gas-phase abundance does not coincide with the one of methanol ice. This is related with high gas and dust density in the disc regions at $r<50$ au leading to a low efficiency of CH$_3$OH ice desorption in comparison with corresponding freeze out timescales. At larger radii, where the disc becomes less dense and more transparent to the interstellar UV radiation and freeze out of gas-phase CH$_3$OH is slower, the CH$_3$OH gas-phase abundance increases and reaches its maximum, leading to the appearance of a ring of methanol emission with a radius of $\sim60$ au. This also explains why the ring of methanol emission in our model does not coincide with the CO snow line located at $\sim30$~au from the TW Hya disc centre \citep{Qi2013}.

The TW~Hya disc physical structure is also responsible for relatively small deviations from LTE in CH$_3$OH transitions. For the strongest CH$_3$OH transitions, including those observed by W16, the critical density is lower than the gas density by a factor of $>10$ throughout most of the disc regions with a relatively high ($X_{\rm{M}}>10^{-12}$) methanol abundance. This is in contrast to the predictions of \citet{Parfenov2016} based on the DM~Tau disc model, according to which LTE and non-LTE CH$_3$OH line flux densities can differ by up to two orders of magnitude. Thermalization of the strongest methanol transitions in the TW~Hya disc makes them a reliable probe of the disc thermal structure. The strong transitions with $F_{\nu}>10$~mJy that show maximum deviations from LTE occur in the frequency range of $500$--$600$~GHz and can not be accessed with ALMA due to a frequency gap between ALMA Bands 8 and 9.

In this study, we have used the TW~Hya disc physical model of \citet{Gorti2011} that is different from the model of \citet{Kama2016}. This difference can be considered as an uncertainty in TW~Hya disc physical parameters. Our calculations have shown that this uncertainty does not affect significantly the results of LRT calculations. We also expect that this uncertainty does not affect the CH$_3$OH distribution in the disc. The methanol line flux densities are much more sensitive to the uncertainties of the disc chemical model.

One of the main parameters of chemical models that influence the abundances of complex species is $E_{\rm{diff}}/E_{\rm{b}}$ \citep[see e.g.][]{Walsh2014}. In this study, we considered the models with $E_{\rm{diff}}/E_{\rm{b}}=0.4$ and 0.77 which are close to the extreme values adopted in chemical studies. The former model is consistent with the observations while the latter model underpredicts the line flux densities by a factor of several compared to the W16 observations.
The influence of other chemical model parameters like the probability of chemical desorption as well as physical parameters like the dust grain size distribution is a subject for future studies. 

\section*{Acknowledgements}
We thank an anonymous referee for providing valuable suggestions to improve this paper. S. Parfenov, A. Sobolev and A. Shapovalova are financially supported by the Russian Science Foundation (project no. 15-12-10017). DAS acknowledges support from the Heidelberg Institute of Theoretical Studies for the project ''Chemical kinetics models and visualization tools: Bridging biology and astronomy''.
This research made use of NASA's Astrophysics Data System. The figures in this paper were constructed with the \textsc{matplotlib} package \citep{Hunter2007} and \textsc{aplpy} package hosted at \url{http://aplpy.github.io}.

%%%%%%%%%%%%%%%%%%%%%%%%%%%%%%%%%%%%%%%%%%%%%%%%%%


\bibliographystyle{mnras}
\bibliography{references}

%%%%%%%%%%%%%%%%%%%%%%%%%%%%%%%%%%%%%%%%%%%%%%%%%%

% Don't change these lines
\bsp	% typesetting comment
\label{lastpage}
\end{document}